\documentclass[conference]{IEEEtran}
\usepackage[sort]{cite}
\usepackage{color}
\usepackage{amssymb}
\usepackage{amsmath}
\usepackage{amsthm}
\usepackage{verbatim}
\usepackage{enumerate}
\usepackage{tikz}

\newtheorem{theorem}{Theorem}

\newtheorem{lemma}{Lemma}

\newtheorem{proposition}{Proposition}

\newtheorem{definition}{Definition}
\newtheorem{remark}{Remark}
\usepackage{array}
\usepackage{amssymb}
\usepackage{amsmath}

\newcommand{\prob}{\mathrm{Prob}}

\newcommand*\lon{%
       \mskip1mu
        \relax
        {:}%
        \mskip1mu
        \relax
}

\author{
\IEEEauthorblockN{%
Tarik Kaced\\ 
\IEEEauthorblockA{%
LIRMM \& Univ. Montpellier 2\\
Email: \texttt{tarik.kaced@lirmm.fr}
}
\and  
 Andrei Romashchenko
 \\
\IEEEauthorblockA{%
LIRMM, CNRS \& Univ. Montpellier 2\\ On leave from IITP, Moscow\\
Email: \texttt{andrei.romashchenko@lirmm.fr}
}
}
}

\title{On the Non-robustness of Essentially Conditional Information Inequalities}

\begin{document}

\maketitle

\begin{abstract}

We show that two essentially conditional linear inequalities for Shannon's
entropies (including the Zhang--Yeung'97 conditional inequality) do not hold
for asymptotically entropic points. This means that these inequalities are
non-robust in a very strong sense. This result raises the question of the
meaning of these inequalities and the validity of their use in
practice-oriented applications.

\end{abstract}

\section{Introduction}

Following Pippenger \cite{pippenger85} we can say that the most basic and
general ``laws of information theory'' can be expressed in the language of
information inequalities (inequalities which hold for the Shannon entropies of
jointly distributed tuples of random variables for every distribution). The
very first examples of information inequalities were proven (and used) in
Shannon's seminal papers in the 1940s.  Some of these inequalities have a clear
intuitive meaning.  For instance, the entropy of a pair of jointly distributed
random variables $a,b$ is not greater than the sum of the entropies of the
marginal distributions, i.e., $H(a,b)\le H(a)+ H(b)$. In standard notations,
this inequality means that the mutual information between $a$ and $b$ is
non-negative, $I(a\lon b)\ge0$; this inequality becomes an equality if and only
if $a$ and $b$ are independent in the usual sense of probability theory. These
properties have a very natural  meaning: a pair cannot contain more
``uncertainty'' than the sum of ``uncertainties'' in both components.  This
basic statement can be easily explained, e.g., in term of standard coding
theorems: the average length of an optimal code for a distribution $(a,b)$ is
not greater than the sum of the average lengths for two separate codes for $a$
and $b$.  Another classic information inequality $I(a\lon b | c)\ge0$  is
slightly more complicated from the mathematical point of view, but is also very
natural and intuitive. Inequalities of this  type are called basic Shannon's
inequality, \cite{yeung-book}.

\begin{comment}
Any convex combination of the basic inequalities mentioned above
(possibly applied to different tuples of random variables) holds for any
distribution. Of course, we cannot say that each of the resulting linear
inequalities has a natural intuitive meaning. Though some of these information
inequalities look very intuitive. Let us mention that, for instance, one 
well-known inequality
 $$
 H(c)\le H(c| a) + H(c | b) + I(a\lon b),
 $$
which claims basically that if $c$ can be easily extracted from $a$ and from
$b$ (both conditional entropies $H(c | a)$ and $H(c| b)$ are small) then the
amount of information in $c$ cannot be greater than the mutual information
between $a$ and $b$. 
(This simple observation can be reformulated, e.g., as a version of coding
theorem for a multi-source network.) 
From a technical point of view this inequality is just a corollary of two basic
inequalities $H(c| a,b)\ge 0$ and $I(a\lon b| c)\ge 0$.
\end{comment}

We believe that the success of Shannon's information theory in a myriad of
applications (in engineering and natural sciences as well as in mathematics and
computer science) is due to the intuitive simplicity and natural 
interpretations of the very basic properties of Shannon's entropy. 

Formally, information inequalities are just a dual description of the set of
all entropy profiles. That is, for every joint distribution of an $n$-tuple of
random variables we have a vector of $2^n-1$ ordered entropies (entropies of
all random variables involved, entropies of all pairs, triples, of quadruples, etc.
in some fixed order). A
vector in $\mathbb{R}^{2^n-1}$ is called \emph{entropic} if it represents
entropy values of some distribution. The fundamental (and probably very
difficult) problem is to describe the set of entropic vectors for all $n$. It
is known, see \cite{zy97}, that for every $n$ the closure of the set of all
entropic vectors is a convex cone in $\mathbb{R}^{2^n-1}$. The points that
belong to this closure are called \emph{asymptotically entropic} or
\emph{asymptotically constructible} vectors, \cite{matus07}, say a.e. vectors
for short.  The class of all linear information inequalities is exactly the
dual cone to the set of a.e. vectors. In \cite{pippenger85}
and \cite{han} a natural question was raised: \emph{What is the class  of all
universal information inequalities?} (Equivalently, how to describe the cone of
a.e. vectors?) More specifically, does there exist any linear information
inequality that cannot be represented as a convex combination of Shannon's 
basic inequality?
 
In 1998 Z.~Zhang and R.W.~Yeung came up  with the first example of a
\emph{non-Shannon-type} information inequality~\cite{zy98}:
$$
I(c\lon d)\le 2 I(c\lon d | a) + I(c\lon d | b) + I(a\lon b) 
 + I(a\lon c | d) + I(a\lon d | c).
$$
This unexpected result raised other challenging questions: \emph{What does this
inequality mean? How to understand it intuitively?} Although we still do not
know a complete and comprehensive answer to the last questions,  we have
several  interpretations and explanations of this inequality.  Some
information-theoretic interpretations were discussed, e.g., in
\cite{zhang03,romash03}. This inequality is closely related to \emph{Ingleton's
inequality} for ranks of linear spaces, \cite{ingleton,matus07,dfz07}.  This
connection was explained by F.~Mat\'{u}\v{s} in his paper \cite{matus07-adh},
where the connection  between information inequalities and polymatroids was
established. Mat\'{u}\v{s} proved  that a polymatroid with the ground set of
cardinality $4$ is selfadhesive if  and only if it satisfies the Zhang--Yeung
inequality formulated above  (more precisely, a polymatroid must satisfy all
possible instances of this   inequality for different permutations of
variables).

Thus, the inequality from~\cite{zy98} has some explanations and intuitive
interpretations.  However, another type of inequalities is still much less
understood.  We mean other ``universal laws of information theory'', those that
can be expressed as  \emph{conditional linear information inequalities} (linear
inequalities for entropies which are true for distributions whose entropies
satisfy some linear constraints; they are also called in the literature
\emph{constrained information inequalities}, see \cite{yeung-book}). 
We do not give a general definition of a
``conditional linear information inequality'' since the entire list of all
known nontrivial inequalities in this class is very short.  Here are three of
them:

\begin{itemize}
\item[(1)] \cite{zy97}: if $I(a\lon b|c)=I(a\lon b)=0$, 
then $$I(c\lon d)\le I(c\lon d| a)+ I(c\lon d| b),$$
\item[(2)]  \cite{matus99}: if $I(a\lon b| c)=I(b\lon d|c)=0$, 
then $$I(c\lon d)\le I(c\lon d| a)+ I(c\lon d| b)+I(a\lon b),$$
\item[(3)] \cite{kr11}: if $I(a\lon b| c)=H(c|a,b)=0$,
then $$I(c\lon d)\le I(c\lon d| a)+ I(c\lon d| b)+I(a\lon b).$$
\end{itemize}
It is  known that (1-3) are ``essentially conditional'', i.e., they cannot be extended 
to any unconditional inequalities, \cite{kr11}, e.g., for (1) this means that for any
values of ``Lagrange multipliers'' $\lambda_1, \lambda_2$ the corresponding
unconditional extension
$$
I(c\lon d)\le  I(c\lon d|a)+ I(c\lon d|b) + 
               \lambda_1I(a\lon b) + \lambda_2I(a\lon b| c)
$$
does not hold for some distributions $(a,b,c,d)$. In other words, (1-3) make
some very special kind of ``information laws'': they cannot be represented as
``shades'' of any unconditional inequalities on the subspace corresponding to
their linear constraints.

A few other nontrivial conditional information inequalities can be obtained from the 
results of F.~Mat\'{u}\v{s} in  \cite{matus99}. For example, Mat\'{u}\v{s} proved
that for every integer $k>0$ and for all $(a,b,c,d)$
\begin{eqnarray}%\label{matus-un}%
\nonumber(*)\
\begin{split}
 I(c\lon d)\le  I(c\lon d|a)+ I(c\lon d|b) + I(a\lon b) +\frac1k I(c\lon d|a)\\
       + \frac{k+1}2(I(a\lon c| d)+I(a\lon d|c))
\end{split}         
\end{eqnarray}              
(this is a special case of  theorem~2 in~\cite{matus99}).
Assume that $I(a\lon c| b)=I(b\lon c|a)=0$. Then, as $k\to\infty$ 
we get from~(*) another conditional inequality:
\begin{itemize}
\item[(4)]  if $I(a\lon c| d)=I(a\lon d|c)=0$,
then $$I(c\lon d)\le I(c\lon d| a)+ I(c\lon d| b)+I(a\lon b).$$
\end{itemize}
%\begin{comment}
It can be proven that (4) is also an essentially conditional inequality, i.e.,
whatever are the coefficients $\lambda_1,\lambda_2$, 
$$
I(c\lon d)\le  I(c\lon d|a)+ I(c\lon d|b) + I(a\lon b)
              + \lambda_1I(a\lon c|d) + \lambda_2I(a\lon d| c)
$$
does not hold for some distribution $(a,b,c,d)$.
%\end{comment}

Since (*) holds for a.e. vectors, (4) is also true  for a.e. vectors.
Inequality~(4) is robust in the following sense. Assume that entropies
of all variables involved are bounded by some $h$. 
Then for every $\varepsilon>0$ there exists a $\delta=\delta(h,\varepsilon)$
such that
\begin{itemize}
\item[] if $I(a\lon c| d) \le \delta$ and $I(a\lon d|c)\le \delta$,
then $$I(c\lon d)\le I(c\lon d| a)+ I(c\lon d| b)+I(a\lon b)+\varepsilon$$
\end{itemize}
(note that $\delta$ is not linear in $\varepsilon$).
%
%Inequalities (1-3) remain quite mysterious, and we do not know any intuitive
%explanation of their meaning.  
In this paper we  prove that this is not the case for (1) and (3) -- these inequalities 
do not hold for a.e. vectors, and they are not robust. 
So, these inequalities are, in some sense, similar to the nonlinear
(piecewise linear) conditional information inequality from
\cite{matuspiecewise}.

Together with \cite{kr11}, where (1--3) are proven to be
essentially conditional, our result indicates that (1) and (3) are very
fragile and non-robust properties of entropies. We cannot hope that similar inequalities
hold when the constraints become soft.  For instance, assuming that $I(a\lon
b)$ and $I(a\lon b| c)$ are ``very small'' we cannot say that $$I(c\lon d)\le
I(c\lon d| a)+ I(c\lon d| b)$$ holds also with only ``a small error''; even a
negligible deviation from the conditions in (1) can result in a dramatic effect
$I(c\lon d)\gg I(c\lon d| a)+ I(c\lon d|b)$.

Conditional information inequalities (in particular, inequality (2)) were used
in \cite{matus99} to describe  conditional independences among several  jointly
distributed random variables. Conditional independence is known to have wide
applications in statistical theory (including methods of parameter
identification, causal inference, data selection mechanisms, etc.), see, e.g.,
surveys in \cite{dawid79,pearl09}. We are not aware of any direct or implicit
\emph{practical} usage of~(1-3), but it would not be surprising to see
such usages in the future. However, our results indicate that these
inequalities are non-robust and therefore might be misleading in
practice-oriented applications.
 
 \begin{comment}
So, (1) and (3) can be used only with the assumption that the corresponding 
independence conditions hold exactly, without any error.
Can we assume that some $a$ and $b$ are \emph{absolutely independent}
(respectively, absolutely independent conditional on $c$) when we deal with
objects in the real world?  We do not try to answer this question.  Our
knowledge is not enough to say whether ``essentially conditional'' information
inequalities are just an artifact of the definition of Shannon's entropy for
discrete distributions, or they still make  some ``physical'' meaning. But
certainly these inequalities must be handled and applied with great caution.
\end{comment}

The rest of the paper is organized as follows. We provide a new proof of
why two conditional inequalities (1) and (3) 
are essentially conditional. This  proof
uses a simple algebraic example of random variables. Then, we show that (1) and (3)
are not valid for a.e. vectors, leaving the question for (2) open.

\section{Why ``essentially conditional'' : an algebraic counterexample}

Consider the quadruple $(a,b,c,d)_q$ of geometric objects, resp. $\mathcal{A},
\mathcal{B}, \mathcal{C}, \mathcal{D},$ on the affine plane over the finite
field $\mathbb{F}_q$ defined as follows :

\begin{itemize}

\item First choose a random non-vertical line $\mathcal{C}$ defined by the
equation $y=c_0 + c_1x$ (the coefficients $c_0$ and $c_1$ are independent
random elements of the field); 

\item pick points $\mathcal{A}$ and
$\mathcal{B}$ on $\mathcal{C}$ independently and uniformly at random
(these points coincide with probability $1/q$);

\item then pick a parabola $\mathcal{D}$ uniformly at random in the set of all
non-degenerate parabolas $y=d_0+d_1x+d_2x^2$ (where
$d_0,d_1,d_2\in\mathbb{F}_q, d_2\neq 0$) that intersect $\mathcal{C}$ at
$\mathcal{A}$ and $\mathcal{B}$; (if $\mathcal{A}=\mathcal{B}$ we require that
$\mathcal{C}$ is a tangent line to $\mathcal{D}$). When $\mathcal{C}$ and
$\mathcal{A},\mathcal{B}$ are chosen, there exist $(q-1)$ different parabolas
$\mathcal{D}$ meeting these conditions.

\end{itemize}

A typical quadruple is represented on Figure~1.%\ref{typquad} :
\begin{figure}[!h]
\centering
\begin{tikzpicture}[scale=1.75]

 \draw[very thick] (0,0) rectangle (4,4);
 \clip (0.1,0.1) rectangle (3.9,3.9);

 \draw[thick] (0,3) parabola bend (2,1) (4,4);
 \draw[thick] (0,1.5) -- (4,2.5);

\node[fill=black,inner sep=1.5pt,label=below left:$\mathcal{A}$,circle] 
      (A) at (0.81,1.70) {};
\node[fill=black,inner sep=1.5pt,label=below right:$\mathcal{B}$,circle] 
      (B) at (3.333,2.33) {};
\node[above] (C) at (2,2.05) {$\mathcal{C}$};
\node[below] (D) at (2.1,0.95) {$\mathcal{D}$};
\end{tikzpicture}
\caption{An algebraic example}\label{typquad}
\end{figure}
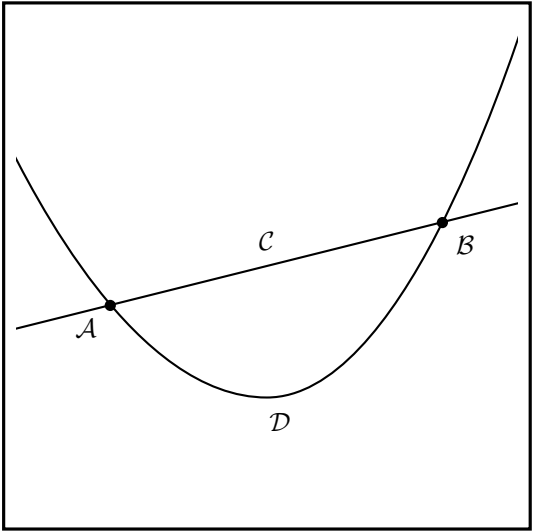

\begin{remark} This picture is not strictly accurate, for the plane is
discrete, but helps grasping the general idea since the relevant properties
used are also valid in the continuous case.  
\end{remark}

Let us now describe the entropy profile of this quadruple.
\begin{itemize}

\item Every single random variable is uniform over its support.

\item 
The line and the parabola share some
mutual information, (the fact that they intersect) which is approximately one
bit. Indeed, $\mathcal{C}$ and $\mathcal{D}$ intersect iff the corresponding
equation discriminant is a quadratic residue, which happens almost half of the
time.
$$I(c\lon d) = \frac{q-1}{q}$$

\item
When an intersection point is given, the line does not give more
information about the parabola.
$$I(c\lon d|a) = I(c\lon d|b) = 0$$

\item
When the line is known, an intersection point does
not help knowing the other (by construction). 
$$I(a\lon b|c) = 0$$

\item
The probability that there is only one intersection point is $1/q$. In
that case, the line can be any line going through this point.
$$I(a\lon b) =  H(c|a,b) = \frac{\log_2 q}{q} $$

\end{itemize}

Now we plug the computations into the following inequalities
$$
I(c\lon d)\le  I(c\lon d| a)+ I(c\lon d| b) 
               + \lambda_1 I(a\lon b) + \lambda_2  I(a\lon b| c) 
$$
or
$$
I(c\lon d)\le  I(c\lon d| a)+ I(c\lon d| b) + I(a\lon b)
               + \lambda_1 I(a\lon b|c) + \lambda_2  H(c| a,b),
$$
which are ``unconditional'' counterparts of (1) and (3) respectively.
For every constants $\lambda_1, \lambda_2$ we get 
$$1 - \frac1{q} \le (\lambda_1+\lambda_2)\frac{\log_2 q}{q} $$
and conclude they can not hold when $q$ is large. 
Thus, we get the following theorem (originally proven in \cite{kr11}):

\begin{theorem}
Inequalities (1) and (3) are essentially conditional.
\end{theorem}

%Should we say that the result is slightly stronger than in [kr11] ? 

\section{Why (1) and (3) do not hold for a.e. vectors}

We are going to use the previous example to show that conditional inequalities
(1) and (3) are not valid for asymptotically entropic vectors. We will use
the Slepian--Wolf coding theorem ({\it cf}. \cite{slepian-wolf}) as our main tool.

\begin{lemma}[Slepian--Wolf] \label{sw-lemma}

Let $(x,y)$ be joint random variables and $(X,Y)$ be $N$ independent copies
of this distribution.
Then there exists $X'$ such that $H(X'|X) = 0$, $H(X') =
H(X|Y) + o(N)$ and $H(X|X',Y) = o(N)$.

\end{lemma}

This lemma constructs a hash of a random variable $X$ which is almost
independent of $Y$ and has approximately the entropy of $X$ given $Y$. We
will say that $X'$ is the Slepian--Wolf hash of $X$ given $Y$ and write $X' =
SW(X|Y)$.

In what follows we call by the \emph{entropy profile} of $(x_1,\ldots,x_n)$  the vector
of entropies for all non-empty subset of these random variable in the lexicographic order.
We denote it
$$\vec{H}(x_1,\ldots,x_n) = (H(\mathcal{S}))_{\emptyset\neq {\mathcal{S}} \subseteq \{x_1,\ldots,x_n\}}.$$
This is a vector in $\mathbb{R}^{2^n-1}$ (dimension is equal to the number of nonempty
subsets in the set of $n$ elements).

\begin{theorem}\label{main-th}
(1) and (3) are not valid for a.e. vectors.
\end{theorem}
\begin{IEEEproof}
For each given inequality, we construct an asymptotically entropic vector which
excludes it. The main step is to ensure, via Slepian--Wolf lemma, that the
constraints are met. 

\paragraph{An a.e. counterexample for  (1)}

\begin{enumerate}[1.]

\item
Start with the quadruple $(a,b,c,d)_q$ from the previous section for some
fixed $q$ to be defined later. Notice that it does not satisfy the constraints.

\item Serialize it: define a new quadruple $(A,B,C,D)$ such that each entropy is
$N$ times greater. $(A,B,C,D)$ is obtained by sampling $N$ times independently
$(a_i,b_i,c_i,d_i)$ according to the distribution $(a,b,c,d)$ and letting,
e.g., $A=(a_1,a_2,\ldots,a_N)$.

\item Apply Slepian--Wolf lemma to get $A'=SW(A|B)$ such that $I(A'\lon
B)=o(N)$, and replace $A$ by $A'$ in the quadruple. The entropy profile of
$(A',B,C,D)$ cannot vary much from the profile of $(A,B,C,D)$. More 
precisely, entropies
for $A',B,C,D$ differ from the corresponding entropies for $A,B,C,D$
by at most $I(A\lon B) + o(N) = O\left(\frac{\log_2 q}{q}N\right)$. 
Notice that $I(A'\lon B|C)=0$ since $A'$
functionally depends on $A$ and $ I(a\lon b|c) = 0$.
%In fact, all entropies remain the same up to $o(N)$ except for $H(A')$ and
%possibly $H(A',D)$.  

\item Scale down the entropy profile of $(A',B,C,D)$ by a factor of $1/N$. This
 operation can be done within a precision of, say, $o(N)$.  Basically, this can
 be done because the set of all a.e. points is convex (see, e.g., \cite{yeung-book})

\item Tend $N$ to infinity to define an a.e. vector. This limit vector is
\textbf{not} an entropic vector.  For this a.e. vector, inequality (1) does not
hold when $q$ is large. Indeed $I(A\lon B)/N$ and $I(A\lon
B|C)/N$ both approaches zero as $N$ tends to infinity. On the other hand, 
for the resulting limit vector, inequality (1) turns into
\[
1+O\left(\frac{\log_2 q}{q}\right) \le 
O\left(\frac{\log_2 q}{q}\right),
\]
which can not hold if $q$ is bigger than some constant.
\end{enumerate}
\medskip

\paragraph{An a.e. counterexample for  (3)} 
We start with another lemma based on the Slepian--Wolf coding theorem.
\begin{lemma}
\label{rel-lemma}
For every distribution $(a,b,c,d)$ and every integer $N$ there exists a
distribution $(A',B',C',D')$ such that 
\begin{itemize}

\item $H(C'|A', B')=o(N)$,

\item The difference between corresponding components of the entropy profile
 $\vec H(A',B',C',D')$ and $N\cdot\vec H(a,b,c,d)$  is
 at most $N\cdot H(c|a,b) + o(N)$.

\end{itemize}
\end{lemma}
\begin{IEEEproof}
First we serialize $(a,b,c,d)$,  i.e.,
we take $M$ i.i.d. copies of the initial distribution. 
The result of this serialization is a distribution $(A,B,C,D)$ whose entropy profile is
the exactly the entropy profile of $(a,b,c,d)$ multiplied by $M$. In particular, 
we have $I(A\lon B|C) = 0$.
Then, we apply Slepian--Wolf encoding (Lemma~\ref{sw-lemma})
and get a $Z=SW(C|A,B)$ such that 
\begin{itemize}
\item $H(Z|C)=0$,
\item $H(Z)=H(C| A,B) + o(M)$,
\item $H(C|A,B,Z)=o(M)$.
\end{itemize}
The entropy profile of the conditional distribution of $(A,B,C,D)$ given $Z$
differs from then entropy profile of $(A,B,C,D)$ by at most $H(Z) = M\cdot
H(c|a,b) + o(M)$.  Also, if in the original distribution $I(a\lon b|c) = 0$,
then $I(A\lon B|C, Z) = I(A\lon B|C) = 0$.

We would like to ``relativize'' $(A,B,C,D)$ conditional on $Z$ and get a new
distribution for a quadruple $(A',B',C',D')$ whose unconditional entropies are
equal to the corresponding entropies of $(A,B,C,D)$ conditional on $Z$.  For different
values of $Z$, the corresponding  conditional distributions on $(A,B,C,D)$ can
be  very different. So there is no well-defined  ``relativization'' of
$(A,B,C,D)$ conditional on $Z$.  The simplest way to overcome this obstacle is
the method  of quasi-uniform distributions suggested by T.H.~Chan and
R.W.~Yeung, see  \cite{chan-yeung}.

\begin{definition}[Quasi-uniform random variables, \cite{chan-yeung}]
A random variable $u$  distributed  on a finite set $\mathcal{U}$ is called
\emph{quasi-uniform} if  the probability distribution function of $u$ is
constant over its support (all values of $u$ have the same probability). That
is, there exists $c>0$ such that  $\prob[u=\mathfrak{u}]\in \{0,c\}$ for all
$\mathfrak{u}\in \mathcal{U}$.
A set of random variables $(x_1,\ldots,x_n)$ is called quasi-uniform if for any
non-empty subset $\{i_1,\ldots,i_s\}\subset\{1,\ldots,n\}$ the joint
distribution $(x_{i_1},\ldots,x_{i_s})$ is quasi-uniform.
\end{definition}

In \cite{chan-yeung}[theorem 3.1] it is proven that for every distribution $(A,B,C,D,Z)$ 
and every $\delta>0$ there exists a quasi-uniform distribution $(A'',B'',C'',D'',Z'')$
and an integer $k$ such that 
\begin{equation*}
\| \vec H(A,B,C,D,Z) - \frac1k \vec H(A'',B'',C'',D'',Z'')\|<\delta.
\end{equation*}
For a quasi-uniform distribution for all values $\mathfrak{z}$ of $Z''$ the
corresponding conditional distributions $(A'',B'',C'',D'')$ have the same
entropies, which are equal to the conditional entropies. That is, entropies of
the distribution of $A''$, $B''$, $(A'', B'')$, etc.  given $Z''=\mathfrak{z}$
are equal to $H(A''|Z'')$, $H(B''|Z'')$, $H(A'', B''|Z'')$ and so on. Thus, for 
a quasi-uniform distribution we can do ``relativization'' as follows.

Fix any value $\mathfrak{z}$ of $Z''$ and take the conditional distribution  on
$(A'',B'',C'',D'')$ given $Z''=\mathfrak{z}$.
In this conditional distribution the entropy of 
$C''$ given $(A'',B'')$ is not greater than
\[ k\cdot (H(C|A,B,Z) + \delta)=k\cdot (\delta +o(M)).\]
Also, by letting $\delta$ be small enough (e.g., $\delta=1/M$), 
all entropies of $(A'',B'',C'',D'')$ given $Z''=\mathfrak{z}$ differ from
the corresponding entropies of $kM \cdot \vec H(a,b,c,d)$  by at most 
$H(Z'') \le kM\cdot H(c|a,b) + o(kM)$.

Moreover, entropies of $(A'',B'')$ given $(C'',Z'')$ are the same as entropies
of $(A'',B'')$ given $C''$, since $Z$ functionally depends on $C$.  If in
the original distribution $I(a\lon b|c) = 0$, then the mutual information
between $A''$ and $B''$ given $(C'',Z'')$ is $o(kM)$.

Denote $N = kM$ and $(A',B',C',D')$ the above-defined conditional distribution
to get the theorem.
\end{IEEEproof}
\medskip
\paragraph{Rest of the proof for (3)}

\begin{enumerate}[1.]

\item Start with the distribution $(a,b,c,d)_q$ for some $q$, to be fixed later,
from the previous section.

\item Apply the ``relativization'' lemma~\ref{rel-lemma} and get
$(A',B',C',D')$ such that $H(C'|A',B')=o(N)$. Lemma~2 guarantees that other
entropies are about $N$ times larger than the corresponding entropies for
$(a,b,c,d)$, possibly with an overhead of size $$O(N\cdot H(c|a,b)) =
O\left(\frac{\log_2 q}{q}N\right).$$ Moreover, since the quadruple $(a,b,c,d)$
satisfies $I(a\lon b|c) = 0$, we also have $I(A'\lon B'|C') = 0$ by
construction of the random variables in Lemma~\ref{rel-lemma}.
 
\item Scale down the entropy profile of $(A',B',C',D')$ by a factor of $1/N$
within a $o(N)$ precision.  

\item Tend $N$ to infinity to get an a.e. vector. Indeed, all entropies from
the previous profile converge when $N$ goes to infinity. Conditions of
inequality (3) are satisfied for $I(A'\lon B'|C')$ and $ H(C'|A', B')$ both
vanish at the limit. Inequality (3) eventually reduces to
$$
1+O\left(\frac{\log_2 q}{q}\right) \le 
O\left(\frac{\log_2 q}{q}\right)
$$
which can not hold for large enough $q$.

\end{enumerate}
\end{IEEEproof}

\begin{remark}
In both cases of the proof we constructed an a.e. vector such that the corresponding 
unconditional inequalities with Lagrange multipliers reduces (as $N\to \infty$) to

\begin{align*}
1+O\left(\frac{\log_2 q}{q}\right) &\le 
O\left(\frac{\log_2 q}{q}\right) + o(\lambda_1+\lambda_2),
\end{align*}
which cannot hold if we choose $q$ appropriately.
\end{remark}

\begin{remark}
Notice that in our proof even one fixed value of $q$ suffices
to prove that (1) and (3) do not hold for a.e. points. The choice of the value of $q$ provides some freedom in
controlling the gap between the lhs and rhs of both inequalities. 
\end{remark}

\medskip 

In fact, we may combine the two above constructions into one to get a single
a.e. vector to prove the previous result.

\begin{proposition}
There exists one a.e. vector which excludes both (1) and (3) simultaneously.
\end{proposition}
\emph{Proof sketch:}
\begin{enumerate}[1.]

\item Generate $(A,B,C,D)$ from $(a,b,c,d)_q$ with entropies $N$ times greater.

\item Construct $A''=SW(A|B)$ and $C'=SW(C|A,B)$ simultaneously (with the same
serialization $(A,B,C,D)$). 

\item Since $A''$ is a Slepian--Wolf hash of $A$ given $B$, we have 
\begin{itemize}
\item $H(C|A'',B) = H(C|A,B) + o(N)$ and 
\item $H(C|A'',B,C') = H(C|A,B,C') + o(N) = o(N)$.
\end{itemize}

\item By inspecting the proof of the Slepian--Wolf theorem we conclude that
$A''$ can be plugged into the argument of Lemma~\ref{rel-lemma} instead of $A$.
The entropy profile of the quadruple $(A',B',C',D')$ thusly obtained from
Lemma~\ref{rel-lemma}  is approximately $N$ times the entropy profile of
$(a,b,c,d)_q$ with a possible overhead of
$$
O(I(A\lon B) + H(C|A,B)) + o(N) = O\left(\frac{\log_2 q}{q}N\right),
$$
and further :
\begin{itemize}
\item $I(A'\lon B'|C') = 0$,
\item $I(A'\lon B') = o(N)$,
\item $H(C'|A',B') = o(N)$.
\end{itemize}

\item Scale  the corresponding entropy profile by a factor $1/N$ and tend $N$
to infinity to define the desired a.e. vector.

\end{enumerate}

\section{Conclusion \& Discussion}

In this paper we discussed the known conditional information inequalities. We
presented a simple algebraic example which provides a new proof that two
conditional information inequalities are essentially conditional (they
cannot be obtained as a direct corollary of any unconditional information
inequality). Then, we prove a stronger result: two  linear conditional information 
inequalities are not valid for \emph{asymptotically entropic} vectors. 

This last result has a counterpart in the Kolmogorov complexity framework. It
is known that unconditional linear information inequalities for Shannon's
entropy can be directly translated into equivalent linear inequalities for
Kolmogorov complexity, \cite{hrsv}. For conditional inequalities the things are
more complicated. Inequalities (1) and (3) could be rephrased in the
Kolmogorov complexity setting;  but the natural counterparts of 
these inequalities prove to be  not valid for Kolmogorov complexity. 
The proof of this fact is very similar to the
argument in Theorem~\ref{main-th} (we need to use Muchik's theorem on conditional
descriptions \cite{muchnik} instead of  the Slepian--Wolf theorem employed in
Shannon's framework).  We skip details for the lack of space.

%\medskip

%\newpage

\textbf{Open problem 1:} Does (2) hold for a.e. vectors?

%\begin{comment}
Every essentially conditional linear inequality for a.e. vectors has an interesting geometric 
interpretation:
it  provides a proof of Mat\'{u}\v{s}' theorem from  \cite{matus-inf}, which claims that
the convex cone of a.e. vectors  for $4$ variables is not polyhedral.
%(more precisely, this cone cannot be represented as a convex combination of
%any finite number of extremal edges). Thus, 
%(i.e., convex combinations of any finite set of linear 
%information inequalities cannot generate all linear information inequalities
%for four variables, \cite{matus-inf}).

\textbf{Open problem 2:} Do (1)  and (3) (that hold for entropic
but not for a.e. vectors) have any geometric or ``physical'' 
meaning? 
%\end{comment}

\section*{Acknowledgements}
The authors are grateful to Alexander Shen for many useful discussions.
This work was supported in part by NAFIT ANR-08-EMER-008-01 grant.

\end{document}